\documentclass[12pt]{article}

\usepackage{amsmath}
\usepackage{graphicx}
\usepackage{amssymb}
\usepackage{color}
\usepackage{bm}
\usepackage{authblk}
\usepackage{epstopdf}

\usepackage{chngcntr}
\usepackage[labelformat=parens]{subfig}
\usepackage{caption}

\usepackage[inline]{enumitem}

\definecolor{light-gray}{gray}{0.7}

\numberwithin{equation}{section}

\begin{document}

\title{Reactive miscible displacement of light oil in porous media}

\author[1]{M. A. Endo Kokubun\footnote{email: max@xal.no}}
\author[2]{N. Khoshnevis Gargar\footnote{email: negar.khoshnevisgargar@deltares.nl}}
\author[3]{D. Marchesin\footnote{email: marchesi@impa.br}}
\author[4]{J. Bruining\footnote{email: j.bruining@tudelt.nl}}

\affil[1]{\small{\it Expert Analytics, Oslo, Norway}}
\affil[2]{\small{\it Deltares, Unit Geo-Engineering, Delft, The Netherlands}}
\affil[3]{\small{\it Instituto Nacional de Matem\'{a}tica Pura e Aplicada (IMPA), Brazil}}
\affil[4]{\small{\it Delft University of Technology, Civil Engineering and Geosciences, The Netherlands}}

\date{}

\maketitle

\begin{abstract}
We develop a theory for the problem of high pressure air injection into deep reservoirs containing light oil.
{Under these conditions}, the injected fluid (oxygen + inert components) is completely miscible with the oil in the reservoir.
Moreover, exothermic reactions  between dissolved oxygen and oil are possible.
We use Koval's model to account for the miscibility of the components, such that the fractional-flow functions resemble the ones from Buckley-Leverett flow.
This allows to decompose the solution of this problem into a series of waves.
We then proceed to obtain full analytical solutions in each wave.
Of particular interest {is the case where} the combustion wave presents a singularity in its internal wave profile.
Evaluation of the variables of the problem at the singular point determines the macroscopic parameters of the wave, i.e., combustion temperature, wave speed and downstream oil fraction.
The waves structure was observed previously for reactive immiscible displacement and we describe it here for the first time for reactive miscible displacement of oil.
We validate the developed theory using numerical simulations.
\end{abstract}

\section{Introduction}
\label{intro}

\begin{table}[ht]
	\centering
	\caption{Nomenclature}
	\label{symbols}
	\begin{tabular}{|l l |l l|}
	\hline\noalign{\smallskip}
	$c_i$               & saturation of $i$                           & $\alpha_o$          & heat capacities ratio \\
	$C_m,C_o$           & heat capacity of rock and oil (J$/$m$^3$K)  & $\theta$            & dimensionless temperature \\	
        $f_i$               & fractional flow function of $i$             & $\lambda$           & thermal conductivity (W$/$m~K)\\
	$Q$                 & heat of reaction (J$/$m$^3$)                & $\mu_i$             & viscosity of $i$ (c~P)\\
	$R$                 & reaction rate (mol$/$m$^3$s)                & $\nu_i$             & stoichiometric coefficient of $i$\\
	$T$                 & temperature (K)                             & $\sigma$            & velocities ratio\\
			    &						  & $\varphi$           & rock porosity \\ 	
\noalign{\smallskip}\hline
	\end{tabular}
\end{table}

Miscible fluid injection is a recovery technique that permits enhancement in oil recovery due to the reduction (or even elimination) of the interfacial tension between oil and the displacing phase \cite{muggeridge2013}, {even if it deteriorates the mobility ratio.}
Usual field applications consider the injection of an inert gas, such as CO$_2$ or N$_2$.
Recently, it was proposed that injection of high-pressure air into deep reservoirs can improve recovery rates due to combustion \cite{gargar2019}.
The numerical simulations conducted by Gargar {\it et al.} \cite{gargar2019} showed that the exothermic reaction between miscible air and oil forms an oil bank in front of the combustion front.
Thus, the combustion front acts as a piston, pushing the oil towards the extraction site and enhancing oil recovery.
High-pressure air injection (HPAI) is a common enhanced oil recovery (EOR) method in immiscible flow conditions \cite{montes2010high,denney201130}.
Traditionally used in the recovery of heavy oils \cite{dabbous1971situ,hardy1972situ,CastanierAUG2003}, this technique was extended to the recovery of light oils in the past decades \cite{hagoort1976,greaves2000improved,clara2000laboratory,chen2013high}.
In the latter case, thermal expansion and gas drive promoted by the oxidation reaction are responsible for enhancing recovery in immiscible flows.
In this paper we analyse the problem of HPAI into deep reservoir containing light oil, thus considering not only the miscible displacement of gas and oil but also the exothermic reaction between them.

When injection occurs in deep reservoirs, the injected fluid can become totally miscible with the oil due to high reservoir pressure (typically above 100 bars).
In this case, the recovery efficiency depends on a series of factors \cite{blackwell1959factors,blackwell1960recovery}, such as longitudinal dispersion, channeling and difference in the viscosities, which can destabilize the displacement process.
If fingers are formed during the displacement, early breakthrough of oxygen can occur, resulting in poor recovery and a safety issue.
The Koval model \cite{koval1963method} simplifies the description of this complicated problem by considering a Buckley-Leverett-type formulation, thus facilitating the analysis of the miscible displacement problem.
Therefore, the Koval model is a suitable tool to formulate the theory for the miscible reactive flow problem.

In this paper we study the wave structure that arises from the reactive miscible displacement resultant from high-pressure air injection in a deep reservoir of light oil.
In one dimension this problem presents an analytical solution for each wave.
In particular, in the combustion wave, where the reaction between oxygen and oil takes place, we show that the heteroclinic orbit connecting the upstream and downstream equilibrium states of the wave must pass through a singular point.
It turns out that the conditions at the singular point determine the macroscopic properties of the combustion wave, i.e., wave speed and combustion temperature, which ultimately determine the recovery efficiency.
Recently, it was shown that such singularity is relevant for the problem of reactive immiscible displacement \cite{mailybaev2011resonance,kokubun2017}.
Here, we identify the emergence of a similar structure for reactive miscible displacement.

It is worth to point that the idea behind using the Koval model is to simplify the theoretical description.
Moreover, a $1D$ description does not capture the formation of fingers, which can have a major impact on recovery efficiency, e.g., due to oxygen breakthrough.
Thus, in Appendix \ref{secApp2d} we also present some numerical simulations performed in $2D$ without resorting to the Koval model, which shows good qualitative agreement with the results presented herein.

\section{Model formulation}
\label{secModel}

We study the miscible flow problem when air is injected
into a porous rock filled with oil at high pressure.
The injected phase consists of oxygen and inert components.
The injected air is miscible in all proportions with the oil due
to the high pressures, i.e., the reservoir pressure is above the minimal miscibility pressure \cite{wang1998}.
This assumption also allows disregard capillary pressure effects \cite{gerritsen2005modeling}.
Then, the mixed phase consists of three components: oil, oxygen and inert
components.
Their saturations are given, respectively, by $c_1,c_2$ and $c_3$, with the constraint
$c_1 + c_2 + c_3 = 1$.
The solvent contains air and (inert) reaction products, with a saturation $c_s =
c_2+c_3$, whereas the oil has saturation $c_1 = 1 - c_s$.
The oil reacts with oxygen to produce inert components according to
\[
\nu_1[\mbox{oil}] + [O_{2}] \rightarrow \nu_3[\mbox{inert components}]
\]
We disregard any volume change due to reactions, temperature expansion and
compositional mixing.

We consider the Koval model \cite{koval1963method} for the miscible displacement, which models the miscibility of the solvents (air and inert components) into oil.
The miscible displacement is incorporated by considering linear relative
permeabilities proportional to their respective saturations.
Hence, the fractional flow function associated with the Koval model resembles
the one used in the Buckley-Leverett model and for the  $i$-component is given by
\begin{equation}
f_i(c_i,T)
=
\frac{c_i/\mu_i}{\sum_{j=1}^3c_j/\mu_j},
\label{eqMOD.frac}
\end{equation}
where $\sum_{i=1}^3 f_i = 1$.

Conservation equations for the oil, oxygen and inert components are thus respectively given by
\begin{align}
\varphi\frac{\partial c_1}{\partial t}
+
\frac{\partial u f_1}{\partial x}
&=
-\nu_1 R,
\label{eqMOD.coil}
\\[5pt]
\varphi\frac{\partial c_2}{\partial t}
+
\frac{\partial u f_2}{\partial x}
&=
-R,
\label{eqMOD.c2}
\\[5pt]
\varphi\frac{\partial c_3}{\partial t}
+
\frac{\partial u f_3}{\partial x}
&=
\nu_3 R,
\label{eqMOD.c3}
\end{align}
where $\varphi$ is the constant rock porosity and $R\geq0$ the exothermic reaction rate.
Summing Eqs. (\ref{eqMOD.coil})--(\ref{eqMOD.c3}) yields the equation
determining the total Darcy velocity
\begin{equation}
 \frac{\partial u}{\partial x}
 =
 (\nu_3 - 1 - \nu_1)R.
 \label{eqMOD.u}
\end{equation}

{On the pore scale,  the temperature of solid rock, oil and miscible gas are approximately equal and  we can} write the heat balance equation as
\begin{equation}
\frac{\partial}{\partial t}(C_{m}\Delta T + \varphi C_{o}\Delta T)
+
\frac{\partial}{\partial x}(C_{o}u \Delta T)
=
\lambda\frac{\partial^2 T}{\partial x^2} + QR.
\label{eqMOD.t}
\end{equation}

The effective viscosity of the solvent is calculated by the fourth-root mixing rule \cite{koval1963method,poling2001properties}
\begin{equation}
\mu_{mix}^{-1/4}
=
0.22~\mu_{air}^{-1/4} + 0.78~\mu_1^{-1/4},
\label{viscosity}
\end{equation}
with $\mu_{air}(T)$ the viscosity of the mixture of oxygen and inert components assumed to be independent of the composition.
Thus, we have $\mu_2=\mu_3=\mu_{mix}$.

\subsection{Dimensionless equations}

In order to make the governing equations dimensionless, we introduce the
ratios
\begin{equation}
\tilde{t} = \frac{t}{t^*}, \ \ \
\tilde{x} = \frac{x}{x^*}, \ \ \
\theta    = \frac{T-T_{ini}}{\Delta T^*}, \ \ \
\tilde{u} = \frac{u}{\varphi v^*},
\label{eq.dim}
\end{equation}
where the characteristic values are given by
\begin{equation}
t^* = \frac{x^*}{v^*}, \ \ \
x^* = \frac{\lambda}{C_m v^*}, \ \ \
v^* = \frac{Q u^{inj}}{C_m \Delta T^*}, \ \ \
\Delta T^* = T^* - T_{ini},
\label{eq.star}
\end{equation}
and $T^*$ is some characteristic temperature.

After introducing the following dimensionless parameters
\begin{equation}
\alpha_o = \frac{\varphi C_o}{C_m}, \ \ \
\sigma   = \frac{\varphi v^*}{u^{inj}},
\label{eq.dim2}
\end{equation}
we obtain the following set of dimensionless equations (we drop the tildes for {reasons of concise notation})
\begin{align}
\frac{\partial c_1}{\partial t}
+
\frac{\partial u f_1}{\partial x}
&=
-\nu_1r,
\label{eq.dcoil}
\\[5pt]
\frac{\partial c_2}{\partial t}
+
\frac{\partial u f_2}{\partial x}
&=
-r,
\label{eq.dcox}
\\[5pt]
\frac{\partial u}{\partial x}
&=
(\nu_3 - 1 - \nu_1) r,
\label{eq.dc}
\\[5pt]
\frac{\partial}{\partial t}
(1+\alpha_o)\theta
+
\frac{\partial}{\partial x}\alpha_o u \theta
&=
\frac{\partial^2\theta}{\partial x^2}
+
\sigma r,
\label{eq.dt}
\end{align}
where $r$ is the dimensionless reaction rate.
Since $\mu_2 = \mu_3 = \mu_{mix}$ and $c_1+c_2+c_3=1$, we can
express the fractional flow functions as
\begin{equation}
f_1(c_1,\theta)
=
\frac{c_1/\mu_1}{c_1/\mu_1 + (1-c_1)/\mu_{mix}},
\ \ \
f_2(c_1,c_2,\theta)
=
\frac{c_2/\mu_{mix}}{1/\mu_{mix} + (1/\mu_1-1/\mu_{mix})c_1}.
\label{eq.frac}
\end{equation}
and we note that $f_1$ does not depend on the fraction of oxygen $c_2$.

Given a set of initial and boundary conditions, the solution of the present
problem is given in terms of {a series of waves} \cite{gargar2019}.
In the next Section we will describe each of these waves.

\section{Series of waves}

\begin{figure}[ht]
\begin{center}
\includegraphics[width=0.5\linewidth]{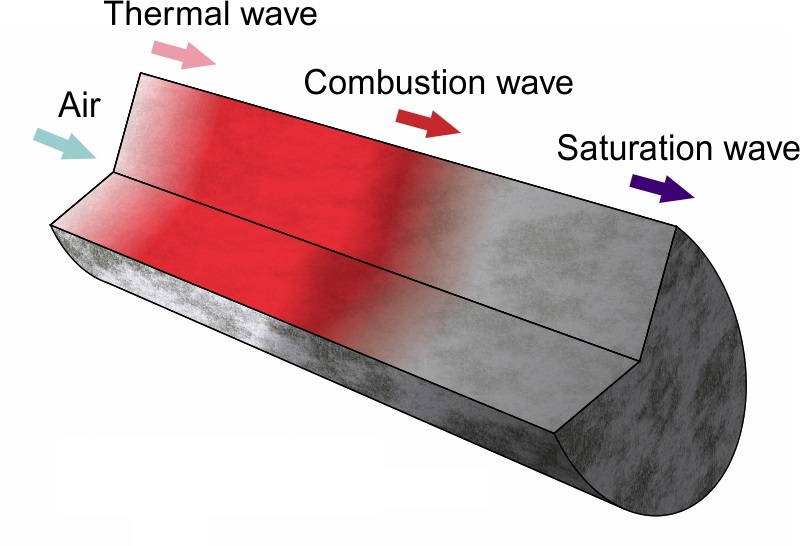}
\caption{Wave sequence: the slower thermal wave, the intermediate combustion wave and the faster saturation wave (figure adapted from \cite{kokubun2017}).}
\label{fig.wave}
\end{center}
\end{figure}

We seek solutions in terms of a series of waves.
The faster, which is located downstream is a rarefaction wave that occurs due to
mixing between oil and inert miscible components.
The slower wave, located upstream, is the thermal wave, where no pure oil is
present, and the liquid is a miscible mixture between air and inerts due to
injection.
In the middle, there is a combustion wave, where the reaction between miscible
oxygen and oil takes place.
A schematic of the wave sequence is shown in Fig. \ref{fig.wave}.

\subsection{Thermal wave}

In the thermal wave, upstream, no reaction takes place due to the absence of oil.
Thus, $r = 0$ in the thermal wave.
Therefore, the energy equation is given by
\begin{equation}
(1+\alpha_o)
\frac{\partial\theta}{\partial t}
+
\alpha_o
\frac{\partial ( u \theta)}{\partial x}
=
\frac{\partial^2\theta}{\partial x^2},
\label{eqTW.dt}
\end{equation}
whereas total mass conservation, Eq. (\ref{eq.dc}) with $r=0$, yields a constant velocity $u = u^{inj} =
\sigma^{-1}$.
Therefore, we have the following equation for the thermal wave
\begin{equation}
\frac{\partial\theta}{\partial t}
+
\frac{\alpha_o}{\sigma(1+\alpha_o)}
\frac{\partial\theta}{\partial x}
=
\frac{1}{(1+\alpha_o)}
\frac{\partial^2\theta}{\partial x^2}.
\label{eqTW.dtB}
\end{equation}
This equation has a well-known solution, describing a wave travelling with speed
$v_T = \alpha_o / (\sigma(1+\alpha_o))$ and broadening proportionally to
$\sqrt{t/(1+\alpha_o)}$.

\subsection{Rarefaction wave}
\label{sec:raref}

The faster wave travels in the reservoir at constant temperature equal to the
initial value $\theta = 0$.
Also, no oxygen is present, $c_2 = 0$, which means that the flow is nonreactive, i.e., $r = 0$.
Therefore, the wave corresponds to the miscible displacement with two components described by the oil concentration $c_1$ and the
concentration of the inert component $c_3 = 1 - c_1$.
The dynamics is governed by an unique equation following from (\ref{eq.dcoil}) as
\begin{equation}
	\frac{\partial c_1}{\partial t}
	+
	u\frac{\partial f_1}{\partial x} = 0,
	\quad
	f_1(c_1,0) =
\frac{c_1}{c_1+(1-c_1)\mu_1/\mu_{mix}},
	\label{R1}
\end{equation}
with constant Darcy speed $u$ (due to incompressibility) and constant
viscosities $\mu_1$ and $\mu_{mix}$.
Equation (\ref{R1}) is the classical Buckley--Leverett equation \cite{Smoller1983};
see also \cite{booth2008miscible} for its application to miscible flows.

In the case of practical interest we have $\mu_1 > \mu_{mix}$, such that the
fractional flow function $f_1$ is convex, i.e., $\partial^2f_1/\partial
c_1^2 > 0$.
Then the profile of a wave with the oil concentration increasing in downstream
direction from the value $c_1^{u}$ to the value $c_1^d$ is described
asymptotically by the rarefaction wave solution (see, e.g., \cite{Smoller1983})
	\begin{equation}
	c_1(x,t) = 
	\left\{
	\begin{array}{ll}
	c_1^u,& x \le uf_1'(c_1^u);
	\\[3pt]
	F(x/t),& uf_1'(c_1^u) < c_1 < uf_1'(c_1^d);
	\\[3pt]
	c_1^d,& x \ge uf'(c_1^d),
	\end{array}
	\right.\\
	\label{R2}
	\end{equation}
{
where $F$, the solution of the Eq. (\ref{R1}) in the rarefaction wave, is a continuous function connecting the constant states $c_1^u$ upstream and $c_1^d$ downstream.
}
This rarefaction wave represents a self-similar profile between two constant states, which
expands linearly with time. 
In order to obtain the solution $F$, we substitute
(\ref{R2}) into (\ref{R1}) {and perform a variable change of the form $\zeta = x/t$}, thus obtaining
\begin{equation}
	\left(u f_1'(F) - \zeta\right)\frac{d F}{d\zeta} = 0.
	\label{eqR2b}
\end{equation}
After dropping the common factor $dF/d\zeta$, this yields
	\begin{equation}
	u f_1'(F) = \zeta.
	\label{R3}
	\end{equation}
For a convex function $f_1$, {i.e., when $\partial^2 f_1\partial c_1^2 > 0$}, this equation provides a unique solution $F(x/t)$ determining the rarefaction wave profile.

%{
%Since oil is consumed in the combustion wave and displaced by the rarefaction wave, we have that $c_1^d > c_1^u$.
%
%Thus, since $f_1''>0$, we have $f_1'(c_1^d) > f_1'(c_1^u)$ and Eq. (\ref{R3}) has a solution $F(\zeta)$ with $uf_1'(c_1^u) < \zeta < uf_1'(c_1^d)$.
%}

\subsection{Combustion wave}

Our main interest lies in the combustion wave, where the exothermic reaction
between miscible oxygen and oil takes place.
As long as the exothermic reaction is confined in the combustion wave, the upstream and downstream states of this wave are equilibrium points of it, i.e., $r=0$.
This holds if the injected oxygen is fully consumed at the reaction point (no leakage) and no oil is left behind by the combustion wave \cite{santos2016,kokubun2016}.

In order to study the combustion wave, we transform to moving coordinates $\xi = x - v t$, where $v$ is the speed of the combustion wave, and obtain the following set of equations
\begin{align}
\frac{d \psi_1}{d\xi}
&= -\nu_1 r,
\label{eq.cwc1}
\\[5pt]
\frac{d \psi_2}{d\xi}
&= - r,
\label{eq.cwc2}
\\[5pt]
\frac{d u}{d\xi}
&= 0,
%(\nu_3-1-\nu_1)r_1,
\label{eq.cwc}
\\[5pt]
\frac{d}{d\xi}
(-v + \alpha_o(u-v))\theta
&=
\frac{d^2 \theta}{d\xi^2}
+
\sigma r,
\label{eq.cwt}
\end{align}
where we defined the fluxes
\begin{equation}
\psi_i = u f_i - v c_i,
\ \ \
i = 1,2,
\label{eq.cwpsi}
\end{equation}
and in Eq. (\ref{eq.cwc}) we considered $\nu_1\ll 1, \nu_3\approx 1$.
Equation (\ref{eq.cwc}) yields a constant velocity in the wave, such that
$u=\sigma^{-1}$.
Note that this means that the total Darcy velocity $u$ does not change across
the three waves.

In the upstream side of the wave, the temperature is at its maximum, no oil is
present, as we consider that it is completely consumed by the reaction, and the
oxygen is at its injection value
\begin{equation}
\xi \rightarrow -\infty: \ \ \
\theta = \theta^u, \ \ \
c_1 = 0, \ \ \
c_2 = c_2^{inj},
\label{eq.upA}
\end{equation}
which yields, for the fluxes
\begin{equation}
\xi \rightarrow -\infty: \ \ \
\psi_1 = 0, \ \ \
\psi_2 = \psi_2^u.
\label{eq.upB}
\end{equation}
In the downstream side of the wave, the temperature is at its initial value, no
oxygen is present and the oil has an unknown saturation $c_1^d$
\begin{equation}
\xi \rightarrow +\infty: \ \ \
\theta = 0, \ \ \
c_1 = c_1^d, \ \ \
c_2 = 0,
\label{eq.downA}
\end{equation}
which yields, for the fluxes
\begin{equation}
\xi \rightarrow +\infty: \ \ \
\psi_1 = \psi_1^d, \ \ \
\psi_2 = 0.
\label{eq.downB}
\end{equation}
The unknowns are $\theta^u, c_1^d$ for the limiting states and the
combustion wave speed $v$.

We can combine Eqs. (\ref{eq.cwc1}), (\ref{eq.cwc2}) and (\ref{eq.cwt}) to
obtain the following reaction-free equations
\begin{align}
\frac{d}{d\xi}
\left(
\psi_1 - \nu_1\psi_2
\right)
&= 0,
\label{eqcw.rfpsi}
\\[5pt]
\frac{d}{d\xi}
\left(
\left(-v + \alpha_o(u-v)\right)\theta
-
\frac{d\theta}{d\xi}
+
\sigma\psi_2
\right)
&=
0.
\label{eqcw.rftA}
\end{align}
These equations can be integrated from upstream to downstream using
(\ref{eq.upA})--(\ref{eq.downB}), yielding
\begin{align}
-\nu_1\psi_2^u
&=
\psi_1^d,
\label{eqcw.psid}
\\[5pt]
\left(
- v(1+\alpha_o) + u\alpha_o
\right)\theta^u
+
\sigma\psi_2^u
&=
0,
\label{eqcw.tu}
\end{align}
Using (\ref{eq.cwpsi}) and (\ref{eq.frac}), Eq. (\ref{eqcw.psid}) is expressed
as
\begin{equation}
\psi_2^u
=
c_2^{inj}(u - v).
\label{eqcw.psidB}
\end{equation}
The upstream (combustion) temperature $\theta^u$ is expressed from Eqs. (\ref{eqcw.tu}) and
(\ref{eqcw.psidB}) as
\begin{equation}
\theta^u
=
\frac{c_2^{inj}(1-\sigma v)}{v(1+\alpha_o) - \alpha_o/\sigma},
\label{eq.tu}
\end{equation}
where we considered $u=\sigma^{-1}$.
Writing Eq. (\ref{eq.tu}) in terms of the speed of the thermal wave $v_T$ yields
\begin{equation}
\theta^u
=
\frac{c_2^{inj}(1-\sigma v)}{(1+\alpha_o)(v - v_T)}.
\label{eq.tuB}
\end{equation}
Since from the ordering of the waves $v>v_T$, a physically  meaningful solution, i.e., $\theta^u>0$, requires
$\sigma v < 1$.

Substituting (\ref{eqcw.psidB}) into Eq. (\ref{eqcw.psid}), using $u = \sigma^{-1}$ and (\ref{eq.cwpsi}) yields
\begin{equation}
f_1(c_1^d,0) = \sigma v c_1^d - \nu_1 c_2^{inj} (1 - \sigma
v).
\label{eqcw.c1d}
\end{equation}

Equations (\ref{eq.tuB}) and (\ref{eqcw.c1d}) determines two of the three
unknowns $\theta^u, c_1^d$ and $v$.
We are missing an equation for the combustion wave speed $v$.
This missing relation is obtained from an analysis of the internal profile of the combustion wave, to be presented in the next Section.

\subsubsection{Internal profile of the combustion wave}
\label{secAppA}

\begin{figure}[ht]
\begin{center}
\includegraphics[width=0.5\linewidth]{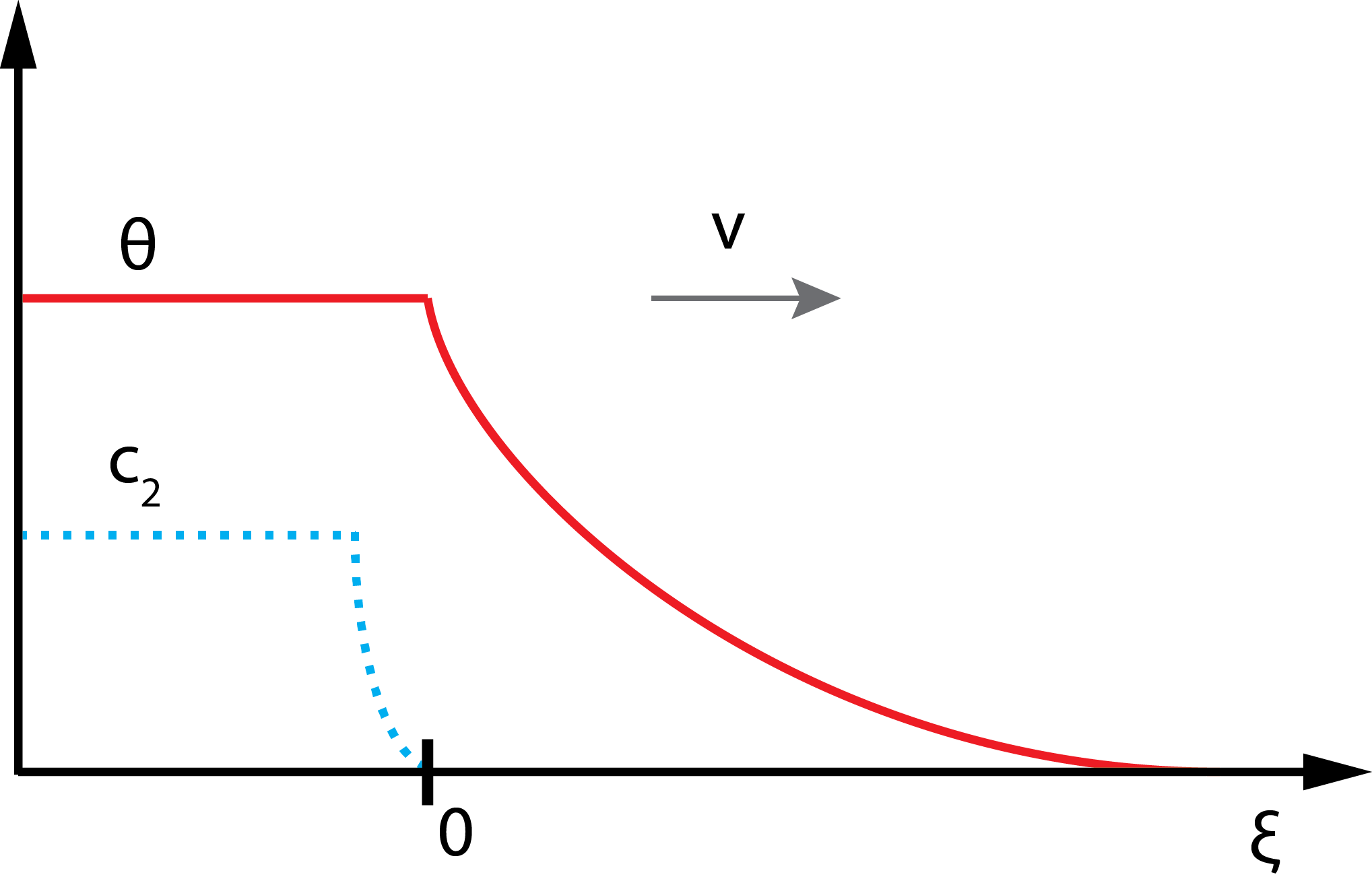}
\caption{Temperature $\theta$ and oxygen fraction $c_2$ profiles along the combustion
wave. Both decrease in the downstream direction of the wave, which travels at a
constant speed $v$. The maximum temperature point is arbitrarily set at $\xi = 0$.}
\label{fig.scheme}
\end{center}
\end{figure}

The missing relation for the combustion wave speed $v$ is obtained from an analysis of the internal profile of the wave.
Equation (\ref{eqcw.rfpsi}) can be integrated from upstream,
$\xi\rightarrow-\infty$, to some internal point of the wave $\xi$, whereas Eq.
(\ref{eqcw.rftA}) can be integrated from downstream, $\xi\rightarrow+\infty$,
to some internal point $\xi$ of the wave, yielding
\begin{align}
\psi_1 - \nu_1 \psi_2
&=
-\nu_1\psi_2^u,
\label{eqI.psi1}
\\[5pt]
\frac{d\theta}{d\xi}
&=
-
\left(v - \frac{\alpha_o}{\sigma}(1-\sigma v)\right)\theta + \sigma\psi_2.
\label{eqI.tA}
\end{align}
Equation (\ref{eqI.psi1}) yields
\begin{equation}
\psi_1 = -\nu_1(\psi_2^u - \psi_2).
\label{eqI.psi12}
\end{equation}
Using the definition of $\psi_1$, we express Eq. (\ref{eqI.psi12}) as
\begin{equation}
u f_1(c_1,\theta) - v c_1 = -\nu_1(\psi_2^u - \psi_2),
\label{eq.surf0}
\end{equation}
where $u=\sigma^{-1}$.
Equation (\ref{eq.surf0}) is a surface in the
$(c_1,c_2,\theta)$-space.

To close the system, we use the equation for $\psi_2$.
Therefore, we have the following system of two differential equations
\begin{align}
\frac{d\psi_2}{d\xi}
&=
-r,
\label{eqI.psi2B}
\\[5pt]
\frac{d\theta}{d\xi}
&=
-
\left(v - \frac{\alpha_o}{\sigma}(1-\sigma v)\right)\theta + \sigma\psi_2,
\label{eqI.tB}
\end{align}
which are valid at the surface $H(c_1,c_2,\theta)=0$ given by
\begin{equation}
H(c_1,c_2,\theta)
=
u f_1(c_1,\theta) - v c_1 + \nu_1(\psi_2^u -
\psi_2),
\label{eq.surf}
\end{equation}
and with an associated vector field given by
\begin{equation}
\frac{d}{d\xi}H(c_1,c_2,\theta) = 0.
\label{eq.surfVec}
\end{equation}

In order to find the singularities of the system, we express the left-hand side
of Eq. (\ref{eqI.psi2B}) and the vector field (\ref{eq.surfVec}) as
\begin{align}
\frac{d\psi_2}{d\xi}
&=
\left(\frac{\partial\psi_2}{\partial c_1}\right)
\frac{dc_1}{d\xi}
+
\left(\frac{\partial\psi_2}{\partial c_2}\right)
\frac{dc_2}{d\xi}
+
\left(\frac{\partial\psi_2}{\partial\theta}\right)
\frac{d\theta}{d\xi},
\label{eq.difpsi2}
\\[5pt]
\frac{d H}{d\xi}
&=
\left(\frac{\partial H}{\partial c_1}\right)
\frac{dc_1}{d\xi}
+
\left(\frac{\partial H}{\partial c_2}\right)
\frac{dc_2}{d\xi}
+
\left(\frac{\partial H}{\partial\theta}\right)
\frac{d\theta}{d\xi}.
\label{eq.difvec}
\end{align}

Therefore, the governing system of equations (\ref{eqI.psi2B}), (\ref{eqI.tB})
and the associated vector field of the surface (\ref{eq.surf}) can be cast into matrix form as
\begin{equation}
\begin{bmatrix}
\partial\psi_2/\partial c_1  &  \partial\psi_2/\partial c_2  &
\partial\psi_2/\partial \theta \\
0                            &  0                            &  1
 \\
\partial H/\partial c_1      &  \partial H/\partial c_2      &  \partial
H/\partial \theta
\end{bmatrix}
\begin{bmatrix}
dc_1/d\xi \\
dc_2/d\xi \\
d\theta/d\xi
\end{bmatrix}
=
\begin{bmatrix}
-r \\
G(c_1,c_2,\theta) \\
0
\end{bmatrix}
\label{eqI.mat}
\end{equation}
where
\begin{equation}
G(c_1,c_2,\theta)
=
-
\left(v - \frac{\alpha_o}{\sigma}(1-\sigma v)\right)\theta + \sigma\psi_2.
\label{eq.g}
\end{equation}

The matrix in the left-hand side of Eq. (\ref{eqI.mat}) is singular if its
determinant is equal zero.
This condition occurs for
\begin{equation}
\frac{\partial\psi_2}{\partial c_2}\frac{\partial H}{\partial c_1}
-
\frac{\partial\psi_2}{\partial c_1}\frac{\partial H}{\partial c_2}
=
0,
\label{eq.surfA}
\end{equation}
upon evaluation yields
\begin{equation}
\left(
u\frac{\partial f_2}{\partial c_2} - v
\right)
\left(
u\frac{\partial f_1}{\partial c_1} - v - \nu_1 u\frac{\partial
f_2}{\partial c_1}
\right)
-
u\frac{\partial f_2}{\partial c_1}
\left(
u\frac{\partial f_1}{\partial c_2} - \nu_1
\left(u\frac{\partial f_2}{\partial c_2} - v\right)
\right)
=
0,
\label{eq.surfB}
\end{equation}
and since $\partial f_1 / \partial c_2 = 0$, (\ref{eq.surfB}) reduces to
\begin{equation}
\left(
u\frac{\partial f_2}{\partial c_2} - v
\right)
\left(
u\frac{\partial f_1}{\partial c_1} - v
\right)
=
0.
\label{eq.foldA}
\end{equation}
Equation (\ref{eq.foldA}) determines the existence of singularities on the
surface defined by (\ref{eq.surf}) in the $(c_1,c_2,\theta)$-space.

We can express the surface $H$ as
\begin{equation}
f_1
=
\frac{v}{u} c_1 - \frac{\nu_1}{u}(\psi_2^u - \psi_2),
\end{equation}
or, using $u=\sigma^{-1}$, the definition of $\psi_2 = u f_2 - v c_2$ and $\psi_2^u$ from (\ref{eqcw.psidB}),
\begin{equation}
f_1(c_1,\theta)
=
\sigma v c_1
-
\nu_1
((c_2^{inj} - f_2) - \sigma v(c_2^{inj}-c_2)).
\label{eq.surfC}
\end{equation}
In terms of $c_1$, the left-hand side of Eq. (\ref{eq.surfC}) is a convex
function, while the right-hand side is approximately a linear function \footnote{The
right-hand side is not strictly linear because $\partial f_2/\partial c_1
\neq 0$.
Nevertheless, since $\nu_1\ll1$, the linear assumption for the right-hand
side of Eq. (\ref{eq.surfC}) is a good approximation.} with a positive
inclination $\sigma v$ and which intersects the vertical axis
at $-\nu_1(1-\sigma v)(c_2^{inj}-c_2)$, because $f_2=c_2$ when
$c_1=0$.
Since $c_2\leq c_2^{inj}$ always holds, the intersection point is less or equal than
zero if $\sigma v < 1$ holds, which is the case for the physically relevant solution $\theta^u>0$, as discussed previously.

Along the combustion wave, from upstream to downstream, both the temperature
$\theta$ and the oxygen fraction $c_2$ decrease - see Fig. \ref{fig.scheme}.
Therefore, the intersection point of the right-hand side of Eq. (\ref{eq.surfC})
at $c_1=0$ decreases from zero to negative values, as we go along the
combustion wave from upstream to downstream.
The inclination remains constant at $\sigma v$.
Since $\partial f_1/\partial\theta>0$, the left-hand side of Eq.
(\ref{eq.surfC}) also decreases as we go from upstream to downstream along the
combustion wave.

\begin{figure}[ht]
\begin{center}
\includegraphics[width=0.5\linewidth]{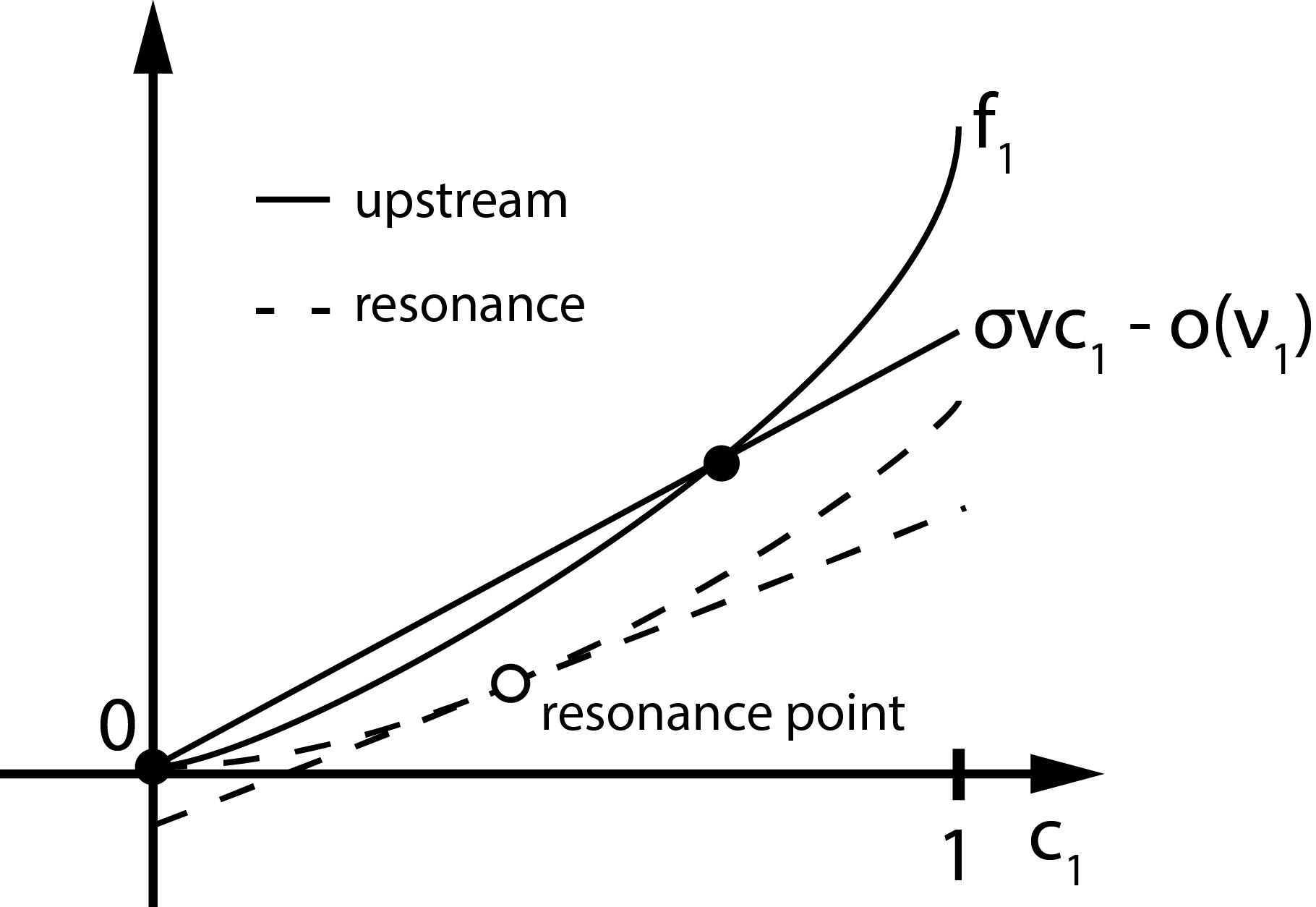}
\caption{Left- and right-hand sides of Eq. (\ref{eq.surfC}) in terms of $c_1$.
Solid lines are conditions in the upstream side, whereas dashed lines represent
conditions at the resonance point (marked as an open circle).}
\label{fig.res}
\end{center}
\end{figure}

The existence of a solution for Eq. (\ref{eq.surfC}) is given by the coincidence
of its left- and right-hand sides.
When $c_1 = 0$, we have that $\partial f_1/\partial c_1 =
\mu_{mix}/\mu_1$.
Therefore, in the upstream side, the solution is unique and at $c_1=0$ if
$\mu_{mix}/\mu_1 \geq \sigma v$, or double-valued (two intersection points)
if $\mu_{mix}/\mu_1 < \sigma v$.
Regardless, the relevant solution is $c_1=0$, as there is no oil in the
upstream side due to complete consumption.
As we go across the combustion wave from upstream to downstream, both left- and
right-hand sides of Eq. (\ref{eq.surfC}) decrease, but $f_1 > 0$ always
holds.
Therefore, a limiting situation is achieved when the coincidence occurs at the
point where their derivatives are equal (as shown in Fig. \ref{fig.res}).
According to (\ref{eq.foldA}), this point is a singular point inside the wave
and which is given by
\begin{equation}
\left.
\left(
u\frac{\partial f_1}{\partial c_1}
\right)
\right|_{res} = v,
\label{eq.sing}
\end{equation}
where the left-hand side of Eq. (\ref{eq.sing}) is evaluated at the singular point, i.e., the resonance point \cite{mailybaev2011resonance,kokubun2017}.

If a travelling wave solution exists and passes through the singular point, it
means that the upstream and downstream states must lie at opposite sides of the
surface (\ref{eq.surfB}) projected into $c_1$.
This can be checked if $\partial H/\partial c_1$ changes sign when
crossing the singular point.
Evaluating $\partial H/\partial c_1$ gives
\begin{equation}
\frac{\partial H}{\partial c_1}
=
u\frac{\partial f_1}{\partial c_1} - v - \nu_1 u \frac{\partial
f_2}{\partial c_1}.
\label{eq.dsurf}
\end{equation}
In the downstream side there is no oxygen, $c_2=0$, such that $\partial
f_2/\partial c_1=0$ and
\begin{equation}
\left.\frac{\partial H}{\partial c_1}\right|_{\xi\rightarrow+\infty}
=
\left.\left(u\frac{\partial f_1}{\partial
c_1}\right)\right|_{\xi\rightarrow+\infty} - v > 0,
\label{eq.dsurfD}
\end{equation}
where the inequality holds because the first term in the right-hand side of Eq.
(\ref{eq.dsurfD}) represents the rarefaction wave, which travels faster than the
combustion wave.

In the upstream side
\begin{equation}
\left.\frac{\partial H}{\partial c_1}\right|_{\xi\rightarrow-\infty}
=
\left.
\left(u\frac{\partial f_1}{\partial c_1}\right)
\right|_{\xi\rightarrow-\infty}
- v - \nu_1 u c_2^{inj}\left(1-\frac{\mu_{mix}}{\mu_1}\right),
\label{eq.dsurfUa}
\end{equation}
or, if we consider $\nu_1\ll 1$,
\begin{equation}
\left.\frac{\partial H}{\partial c_1}\right|_{\xi\rightarrow-\infty}
\approx
\left.
\left(u\frac{\partial f_1}{\partial c_1}\right)
\right|_{\xi\rightarrow-\infty}
- v
=
\left.
\frac{1}{\sigma}
\left(\frac{\mu_{mix}}{\mu_1}\right)
\right|_{\xi\rightarrow-\infty}
-
v.
\label{eq.dsurfUb}
\end{equation}
In order that the travelling wave solution passes through the singularity in the
surface $H$, we must have $\left.\partial H/\partial c_1\right|_{-\infty} <
0$.
Therefore, the following condition must hold
\begin{equation}
\sigma v > \left.\frac{\mu_{mix}}{\mu_1}\right|_{\xi\rightarrow-\infty}.
\label{eq.cond}
\end{equation}
Note that this condition is necessary for the existence of two possible
solutions in the upstream side of the combustion wave, as shown in Fig.
\ref{fig.res}.

At the resonance point, the fraction of oil is an unknown $c_1^r$,
determined from Eq. (\ref{eq.surf0}) evaluated with $c_2=0$ and
$\theta=\theta^u$.
Therefore, we have the necessary conditions to obtain the unknowns of the
problem $\theta^u, v$ and $c_1^d$.

\subsection{Model summary}

From the analysis of the internal profile of the combustion wave performed in the previous Section, it is revealed that the expression for $v$ is obtained by evaluating the following equations
\begin{align}
 u f_1(c_1^r,\theta^u) - v c_1^r
 &=
 - \nu_1(\psi_2(0,c_2^{inj},\theta^u) - \psi_2(c_1^r,0,\theta^u)),
 \label{eq.res1}
 \\[5pt]
 v
 &=
 \left.\left(u\frac{\partial f_1}{\partial c_1}\right)\right|_{c_1^r,0,\theta^u},
 \label{eq.resV}
\end{align}
at a specific point of the combustion wave, the resonance point \cite{mailybaev2011resonance,MTO2013,kokubun2017}, where $c_1=c_1^r$, $c_2=0$ and $\theta=\theta^u$.

Thus, we have all the necessary ingredients to evaluate the macroscopic wave parameters of interest: combustion wave speed $v$, combustion temperature $\theta^u$ and downstream oil saturation $c_1^d$.
In summary, we solve the following set of equations
\begin{align}
 u f_1(c_1^r,0.\theta^u) - v c_1^r
 &=
 - \nu_1(\psi_2(0,c_2^{inj},\theta^u) - \psi_2(c_1^r,0,\theta^u)),
 \label{eq.res1b}
 \\[5pt]
 v
 &=
 \left.\left(u\frac{\partial f_1}{\partial c_1}\right)\right|_{c_1^r,0,\theta^u},
 \label{eq.resVb}
 \\[5pt]
 \theta^u
 &=
 \frac{c_2^{inj}(1-\sigma v)}{v(1+\alpha_o) - \alpha_o/\sigma},
 \label{eq.tub}
 \\[5pt]
 f_1(c_1^d,0)
 &=
 \sigma v c_1^d - \nu_1 c_2^{inj} (1 - \sigma v).
 \label{eqcw.c1db}
\end{align}
Equations (\ref{eq.res1b}) and (\ref{eq.resVb}) evaluated at the internal, resonance point, determines $c_1^r$ and $v$ in terms of $\theta^u$.
Then, these results are used to evaluate $\theta^u$ and $c_1^d$ from Eqs. (\ref{eq.tub}) and (\ref{eqcw.c1db}).
In particular, the values of combustion wave speed and downstream saturation, $v$ and $c_1^d$, respectively, are used to evaluate the rate of oil recovery due to combustion.

In the next Section we analyse the results from our model.

\section{Results}

\begin{table}[ht]
\centering
\begin{tabular}{|l|l|}
\hline
$A_r = 4060$ 1/s                  & $T^* = 600$\,K\\
$C_m = 2$ MJ/m$^{3}$K             & $u^{inj} = 8.0 \times 10^{-7}$ m/s\\
$C_o = 6.7\times10^ 5$ J/m$^{3}$K & $\lambda = 3$ W/m\,K\\
$Q = 13.3$ MJ/m$^ 3$              & $\nu_1 = 0.090$ [mol/mol]\\
$T_{ac} = 7066$\,K                & $\nu_3 = 1.36$ [mol/mol]\\
$T_{ini} = 300$\,K                & $\varphi = 0.3$\\
\hline
\end{tabular}
\caption{Values of reservoir parameters for heptane as a model oil. }
\label{tab1}
\end{table}

In order to validate the theory developed in the previous Section, we present a comparison of our model outputs with numerical solutions.
For such, we consider a dimensional reaction rate of the Arrhenius type in Eqs. (\ref{eqMOD.coil})--(\ref{eqMOD.t}), given by
\begin{equation}
 R = \varphi \rho A_r c_1 c_2~\mbox{exp}\left(-\frac{T_{ac}}{T}\right),
 \label{eq.reac}
\end{equation}
where $A_r$ is the frequency factor, $T_{ac}$ the activation energy and $\rho = 3000~mol/m^ 3$ the molar density of the oil.
Moreover, we consider the following initial and boundary conditions for Eqs. (\ref{eqMOD.coil})--(\ref{eqMOD.t})
\begin{align}
 t &= 0, \ x\geq0: \ \ \ c_1 = 1, \ c_2 = 0, \ \ \ T=T_{ini},
 \\[5pt]
 t &> 0, \    x=0: \ \ \ c_1 = 0, \ c_2 = c_2^ {inj}, \ T = T_{ini}, \ u = u^ {inj},
 \label{eq.icbc}
\end{align}
with $T_{ini}$ the initial temperature, $c_2^ {inj} = 0.21$ the fraction of injected oxygen and $u^{inj}$ the injection velocity.

\begin{figure}[ht]
 \centering
 \includegraphics[width=1\linewidth]{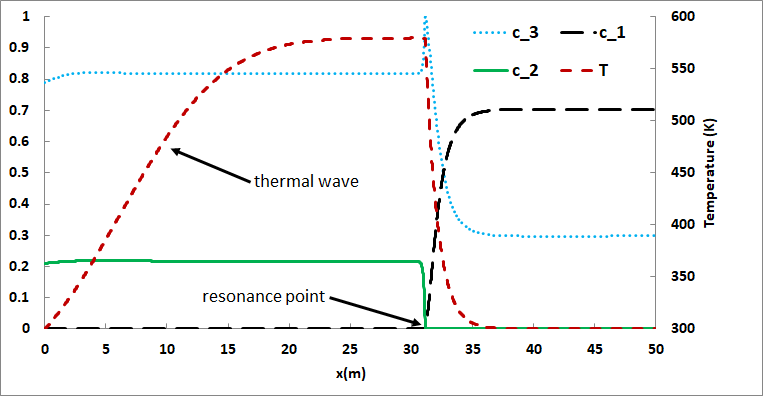}
 \caption{Characteristic combustion wave profile, showing the temperature $T$ and the fractions of oil $c_1$, oxygen $c_2$ and inert components $c_3$. The faster rarefaction wave is not shown, as it already travelled away from the domain. The resonance point and thermal wave are indicated by arrows and the wave speed is $v = 1.6\times10^ {-6}m/s$.}
 \label{fig.numProfile}
\end{figure}

The viscosities of oil and air, necessary for the evaluation of the fourth-root mixing rule (\ref{viscosity}), are given by Sutherland's formula\cite{sutherland1893}  and the Arrhenius model \cite{koval1963method,poling2001properties}
\begin{equation}
 \mu_1(T) = \mbox{exp}\left(\frac{1335.8}{T} - 4.6329\right), \ \ \
 \mu_{air}(T) = \frac{7.5}{T + 120}\left(\frac{T}{291}\right)^{3/2},
 \label{eq.visc}
\end{equation}
and are given in $cP$, with the temperature $T$ in $K$.
The dimensional parameters, necessary for the simulations, are shown in Tab. \ref{tab1} and are given for heptane as a model oil, with $T^* = 600~K$ chosen as the arbitrary characteristic temperature.
The governing equations (\ref{eqMOD.coil})--(\ref{eqMOD.t}) are numerically solved through COMSOL, therefore using a standard Galerkin finite element method with fifth-order Lagrangian polynomial elements.
The domain have spatial length of $L=50~m$, which is enough to capture the formation of the travelling waves, with a grid size of $0.01~m$, fine enough to capture the multi-scale processes.

\begin{figure}[ht]
\centering
  \subfloat[]{\includegraphics[width=0.5\linewidth]{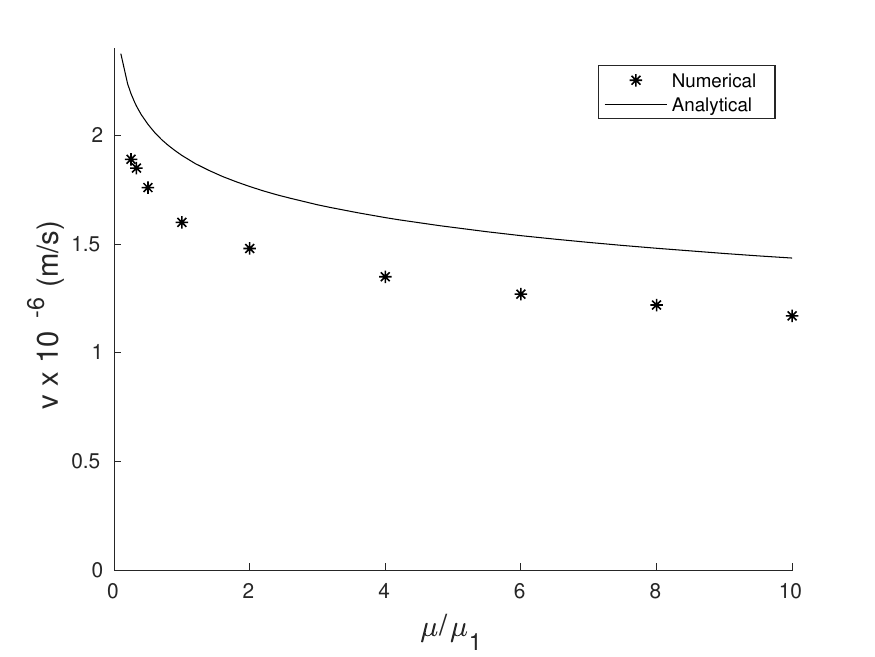}}\hfil
  \subfloat[]{\includegraphics[width=0.5\linewidth]{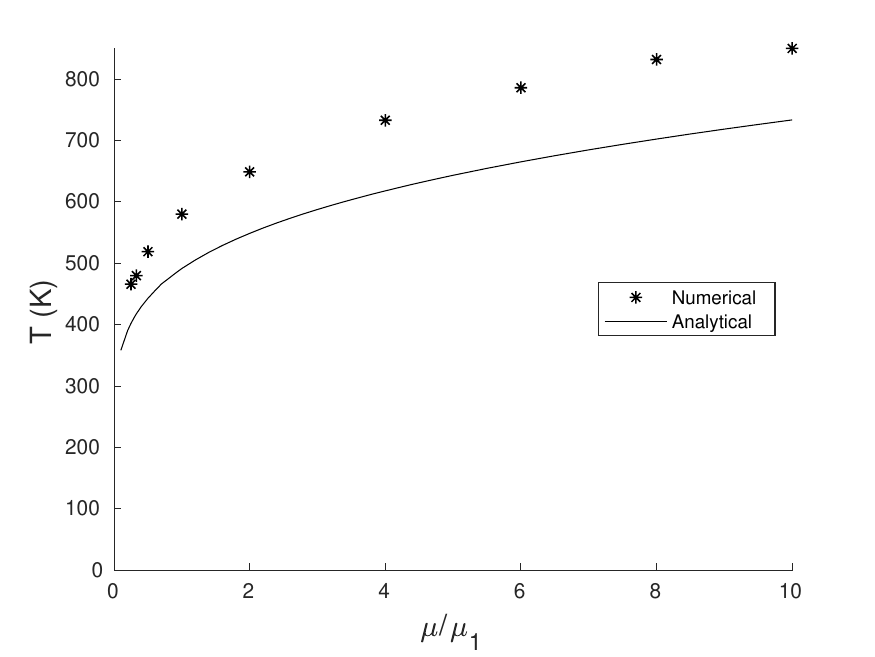}}\\
  \subfloat[]{\includegraphics[width=0.5\linewidth]{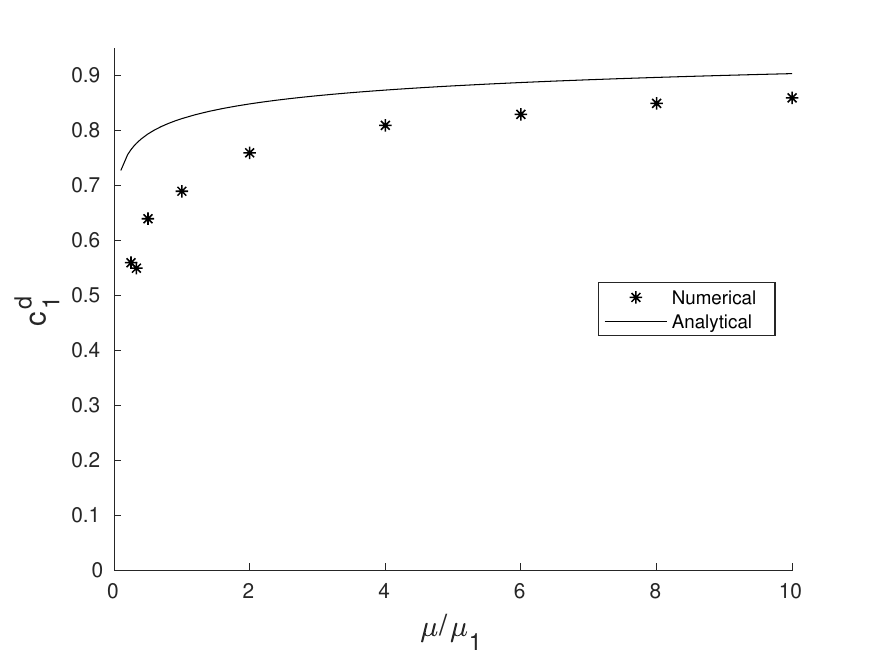}}
  \medskip
  \caption{Comparison between numerical $'*'$ and analytical $'-'$ results for (a) wave speed, (b) combustion temperature and (c) downstream oil fraction. As the viscosity increases, the wave speed decreases, allowing for a longer contact time between oxygen and oil, resulting in higher combustion temperatures and a more efficient recovery. {The discrepancies between analytical and numerical results arise from the approximations $\nu_1 \ll 1$ and $\nu_3\approx1$.}}
  \label{fig.comp}
\end{figure}

A characteristic combustion wave profile, obtained numerically, is shown in Fig. \ref{fig.numProfile}, where we present dimensional profiles for temperature $T$, oil fraction $c_1$, oxygen fraction $c_2$ and inert components fraction $c_3$.
The injection of air occurs at the left and the wave travels towards the right end of the domain.
The reaction is fast due to the large value of the activation energy $T_{ac}$, and which prevents leakage of oxygen and oil through the front.
Oxygen and oil are consumed at $x\approx31~m$, elevating the temperature to $T\approx580~K$ due to the exothermic reaction.
The inert components present a peak at the same point due to the production of combustion products.
The oil fraction at the downstream side of the wave is at a constant state $c_1^d\approx0.7$ and the temperature is at its initial value.
An oil bank is formed at the front of the reaction point and is pushed downstream by the combustion wave in a piston-like displacement, which enhances recovery.
The numerical simulation confirms the existence of the series of waves structure proposed in the derivation of the theory.

In order to validate the theory described in the previous Section, we calculate the values of wave speed $v$, combustion temperature $T^u$ and downstream oil fraction $c_1^d$ for varying values of the oil viscosity $\mu_1$.
We then compare outputs from our theory with numerically-obtained values.
The considered dimensional values yield the dimensionless parameters $\alpha_o = 0.1008$ and $\sigma = 7.333$.
The varying viscosities are given in fractions of the base viscosity $\mu_1$ given by (\ref{eq.visc}).
These results are shown in Figs \ref{fig.comp}, where we compare analytical and numerical results.
As the viscosity of the oil increases, the combustion wave travels at a lower speed, which allows for a longer contact time between oxygen and oil, resulting in higher combustion temperatures.
In turn, this results in a more efficient recovery process, i.e., the downstream oil fraction increases.
Moreover, we see that the trend for increasing viscosity is well captured by the theory.
The maximum errors ($14\%$ for the combustion temperature, $23\%$ for the wave speed and $30\%$ for the downstream saturation) are resultant from the approximations $\nu_1 \ll 1$ and $\nu_3\approx1$ and are within the expected discrepancy, thus validating the theory developed in the last Section.

\subsection{Production profiles}

A fundamental aspect for a successful oil recovery technique is its ability to enhance recovery.
Thus, we compare production curves for the reactive and non-reactive cases.
For the reactive case, the amount of oil recovered at the outlet at a time $t$ is given by $\varphi v c_1^d A t$, where $A$ is the area through where oil is recovered and with $v$ given in its dimensional form.
Therefore, dimensionless production is given by
\begin{equation}
PR_r
=
\frac{\varphi v c_1^d A t}{\mbox{oil}_{ini}},
\label{eq.oilProd}
\end{equation}
where $\mbox{oil}_{ini} = \varphi c_1^{ini} V$ is the initial oil in place, with $V$ the volume of the reservoir.
Since the combustion wave moves at a constant speed $v$, the recovery rate due to combustion is linear in time.

For the non-reactive case, the displacement of oil is given by the reaction-free equation
\begin{equation}
    \frac{\partial c_1}{\partial t} + u\frac{\partial f_1}{\partial x} = 0,
    \label{eq.bl}
\end{equation}
which can be written as
\begin{equation}
    \frac{\partial c_1}{\partial t} + \left(u\frac{\partial f_1}{\partial c_1}\right)\frac{\partial c_1}{\partial x} = 0,
    \label{eq.blb}
\end{equation}
and we recognise $u\partial f_1/\partial c_1$ as the characteristic speed of a point of constant oil saturation.
In immiscible flows, the fractional function $f_1$ has a characteristic $S$-shape.
For those cases, a Buckley-Leverett analysis reveals that the oil is pushed by a shock front which emerges from the injected phase \cite{buckley1942,welge1952}.
In the present case of miscible flow, the fractional flow function $f_1$ is convex, with $\partial f_1/\partial c_1>0$ everywhere, see Fig. \ref{fig.res}.
The description in this case resembles the one for the rarefaction wave described in Section \ref{sec:raref}, but with the downstream state given by $c_1^d = c_1^{ini}$ and the upstream states decreasing from $c_1 = c_1^{ini}$ to $c_1 = 0$.
The production rate of oil for the non-reactive case is thus given by
\begin{equation}
    PR_{nr} = \frac{\varphi c_1^{ini} A t}{\mbox{oil}_{ini}}\left.\left(\frac{\partial f_1}{\partial c_1}\right)\right|_{c_1^u}.
    \label{eq.recNR}
\end{equation}
The velocity of the rarefaction wave decreases as the injected phase moves further into the reservoir, as in this case $c_1$ decreases and $\partial f_1/\partial c_1 > 0$.
Thus, the recovery rate in the non-reactive case is sublinear in time.

\begin{figure}[ht]
\centering
  \subfloat[]{\includegraphics[width=0.5\linewidth]{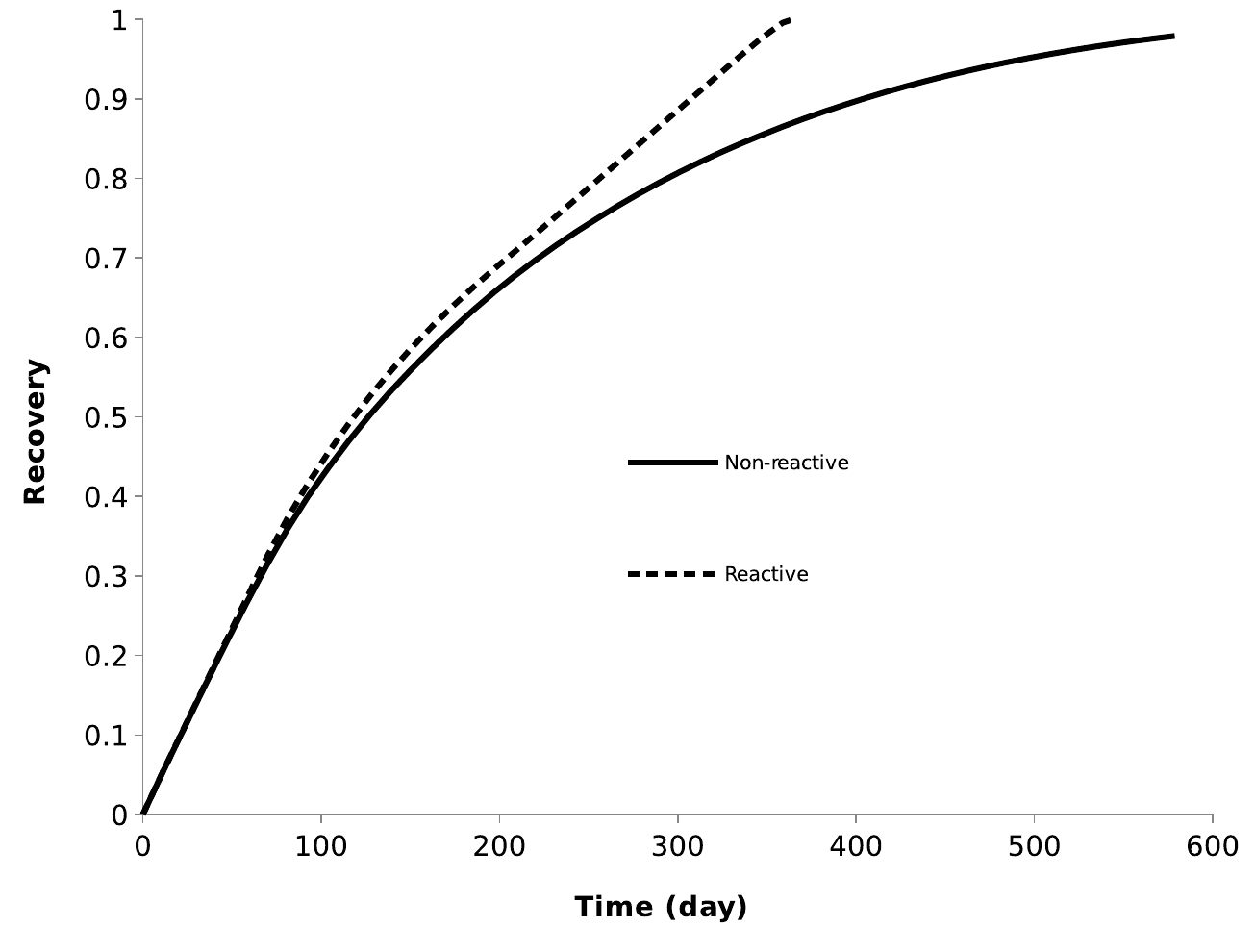}}\hfil
  \subfloat[]{\includegraphics[width=0.5\linewidth]{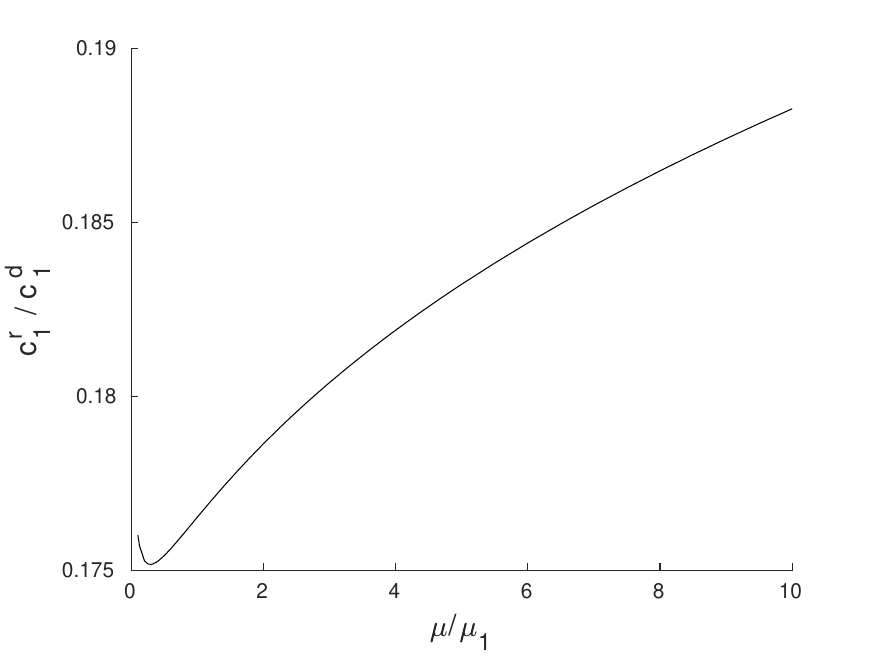}}\medskip
  \caption{(a) Production curve of the fraction of initial oil in place due to combustion (dashed line) compared to the non-reactive case (solid line). (b) Ratio of burned oil to recovered oil as a function of the oil viscosity. The amount of oil burned in comparison to the oil recovered always lies below $19\%$, assuring combustion as a displacement method with a small loss of oil.}
  \label{fig.oilBurned}
\end{figure}

These features can be observed in Fig. \ref{fig.oilBurned}(a), where the oil recovery in terms of initial oil in place is shown for both cases of non-reactive and reactive.
The recovery rates were obtained from the numerical simulations for the oil with viscosity given by (\ref{eq.visc}).
The results from numerical simulations confirm the linear and sublinear natures of the oil recovery process for the reactive and non-reactive cases.
Thus, combustion enhances recovery for miscible flows.

The fast exothermic reaction occurs at the resonance point.
Thus, $c_1^r$ represents the amount of oil burned in the combustion wave.
In Fig. \ref{fig.oilBurned}(b) we present the ratio $c_1^r/c_1^d$  for different oil viscosities.
The ratio $c_1^r/c_1^d$ represents the fraction of the amount of oil burned by the amount of oil recovered, and is therefore a measurement of the viability of this EOR method.
The amount of oil burned in comparison with the oil recovered lies lower than $19\%$ for a $10$-fold increase in the oil viscosity, which shows that the amount of oil burned remains small in comparison with the oil recovered, which is a desirable characteristic for in-situ combustion.
The decrease in the ratio $c_1^r/c_1^d$ when the viscosity increases for low viscosities shown in Fig. \ref{fig.oilBurned}(b) results from the sharp increase in $c_1^d$ with the viscosity, as seen in Fig. \ref{fig.comp}(c).

\section{Conclusions}

We developed a theory for the reactive miscible displacement of oxygen and oil in high pressure reservoirs.
By considering the Koval model for the miscible displacement, we are able to describe the solution of this problem in a series of waves (thermal, combustion and saturation).
The combustion wave, where the exothermic reaction between oxygen and oil takes place, presents a singularity in its internal profile.
The conditions at the singularity determine the macroscopic wave parameters, i.e., wave speed and temperature and is thus ultimately responsible for the recovery efficiency.
The results from the theory are validated with numerical simulations.
Recovery is enhanced by combustion when compared to recovery for the non-reactive case.
For the case with combustion, the recovery rate is linear in time, whereas for the non-reactive case, recovery is sublinear.
The small amount of oil burned with respect to the amount of oil recovered makes this method attractive to enhance recovery of light oil in deep reservoirs.

A singularity in the internal profile of the combustion wave emerges for different reactive displacement mechanisms, i.e., low temperature oxidation \cite{mailybaev2011resonance} and medium temperature oxidation \cite{MTO2013}.
Moreover, such singularity exists even when a multicomponent oil is considered \cite{kokubun2016,kokubun2017}.
A generalized theory for singular wave profiles in a system of balance laws is not yet available, though.
In this paper we present another example of a system possessing this singular structure, i.e., reactive miscible displacement in porous media.

\section*{Acknowledgements}

The work of MAEK was supported by Equinor through the Akademia agreement during the period when the author was a postdoctoral researcher at the University of Bergen.
The work of DM was supported by FAPERJ through the projects PRONEX, CNE and PensaRio (E-26/210.874/2014), and by CAPES through the project NUFFIC.
The Authors greatly acknowledge Prof. Alexei Mailybaev (IMPA) for the original idea and fruitful discussions concerning this work. 
AM work was supported by FAPERJ through the project PensaRio (E-26/210.874/2014)

\appendix

\section{Numerical simulations in $2D$}
\label{secApp2d}

We perform numerical simulations in $2D$ in order to obtain a qualitative comparison with the outputs of our model.
For such, we consider the following set of equations for oil, oxygen and inert components
\begin{align}
    \varphi\frac{\partial c_1}{\partial t}
    +
    \nabla\cdot\left(\bm{u}c_1 - \varphi D\nabla c_1\right)
    &=
    -\nu_1 R,
    \label{eqApp.c1}
    \\[5pt]
    \varphi\frac{\partial c_2}{\partial t}
    +
    \nabla\cdot\left(\bm{u}c_2 - \varphi D\nabla c_2\right)
    &=
    -R,
    \label{eqApp.c2}
    \\[5pt]
    \varphi\frac{\partial c_3}{\partial t}
    +
    \nabla\cdot\left(\bm{u}c_3 - \varphi D\nabla c_3\right)
    &=
    \nu_3 R,
    \label{eqApp.c3}
\end{align}
where we consider molecular diffusion for numerical reasons.
    Summing Eqs. (\ref{eqApp.c1})--(\ref{eqApp.c3}) yields the total mass conservation as
\begin{equation}
    \nabla\cdot\bm{u}
    =
    (\nu_3 - \nu_1 - 1)R.
    \label{eqApp.u}
\end{equation}
The energy equation is given by
\begin{equation}
    (C_m + \varphi C_o)\frac{\partial T}{\partial t}
    +
    \nabla\cdot(C_o\bm{u}T)
    =
    \nabla\cdot(\lambda\nabla T)
    +
    Q R.
    \label{eqApp.t}
\end{equation}
Equation (\ref{eqApp.u}) determines the pressure field through Darcy's law
\begin{equation}
    \bm{u} = - \frac{K}{\mu_{mix}}\nabla p,
    \label{eqApp.p}
\end{equation}
where $\mu_{mix}$, the viscosity of the oil+air mixture, is now determined by
\begin{equation}
    \frac{1}{\mu_{mix}^{1/4}}
    =
    \frac{c_1}{\mu_1^{1/4}}
    +
    \frac{c_2}{\mu_{air}^{1/4}}
    +
    \frac{1-c_1-c_2}{\mu_{air}^{1/4}}.
    \label{eqApp.mu}
\end{equation}

The model presented by Eqs. (\ref{eqApp.c1})--(\ref{eqApp.p}) is a generalisation of the Koval model considered previously.
The main difference lies in the assumption of fractional flow functions and viscosities.

\begin{figure}[ht]
\centering
  \subfloat[]{\includegraphics[width=0.5\linewidth]{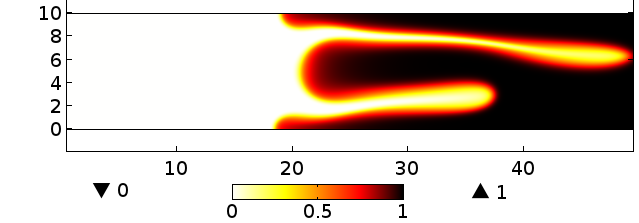}}\hfil
  \subfloat[]{\includegraphics[width=0.5\linewidth]{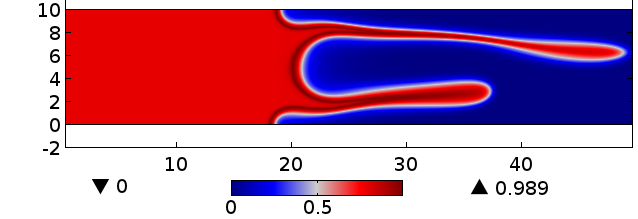}}\\
  \subfloat[]{\includegraphics[width=0.5\linewidth]{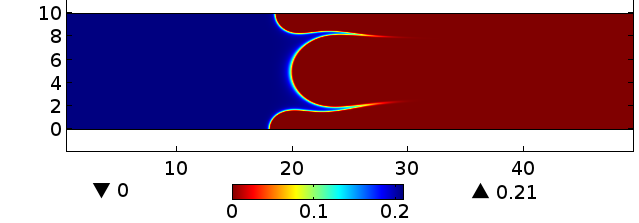}}
  \medskip
  \caption{Results for $2D$ simulations for fractions of (a) oil, (b) inert and (c) oxygen. The inert components finger into the downstream direction, but the injected oxygen is fully consumed by the reaction and no breakthrough occurs. Results are shown for $t=1.87\times10^8~s$ and the flow is from left to right.}
  \label{figApp.comp}
\end{figure}

We consider a $2D$ rectangular domain, with length $L=50~m$ and height $H=10~m$.
The initial and boundary conditions are given by
\begin{align}
 t &= 0, \ x\geq0: \ \ \ c_1 = 1, \ c_2 = 0, \ \ \ T=T_{ini}, \ \ p = 1.00\times10^6~Pa
 \\[5pt]
 t &> 0, \    x=0: \ \ \ c_1 = 0, \ c_2 = 0.21, \ (C_o u_x T-\lambda\partial T/\partial x) = 0, \ \ p = 1.01\times10^6~Pa.
 \label{eqApp.icbc}
\end{align}
For the upper $y=10~m$ and lower $y=0~m$ parts of the domain, we consider no flux boundary conditions for all variables.
The slight difference in the boundary conditions considered here and in the $1D$ case, i.e., for the temperature and pressure, is solely for numerical reasons.
Since in this Section we are only interested in a qualitative comparison, this is not an issue.

\begin{figure}[ht]
\centering
  \subfloat[]{\includegraphics[width=0.5\linewidth]{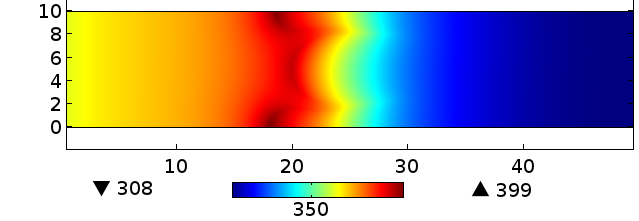}}\hfil
  \subfloat[]{\includegraphics[width=0.5\linewidth]{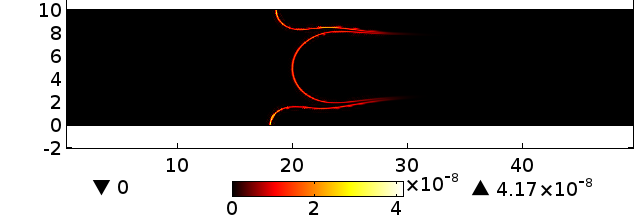}}\medskip
  \caption{Surface plots for (a) temperature and (b) reaction rate $R$, as given by Eq. (\ref{eq.reac}). Results are shown for $t=1.87\times10^8~s$.}
  \label{figApp.temp}
\end{figure}

For the parameters, we consider the values given in Table \ref{tab1}, with the exception of the frequency factor, where we consider a value of $A_r = 6090~1/s$, or a value $50\%$ higher.
This is purely for numerical reasons: to establish a combustion state before the end of the domain.
Note that our theory only requires that oxygen and oil are completely consumed by the exothermic reaction.
Thus, the frequency factor only controls the time scale at which reaction occurs.
Additionally, we consider $D=2\times10^{-9}~m^2/s$ for the value of molecular diffusion.

\begin{figure}[h!]
\centering
  \subfloat[]{\includegraphics[width=0.5\linewidth]{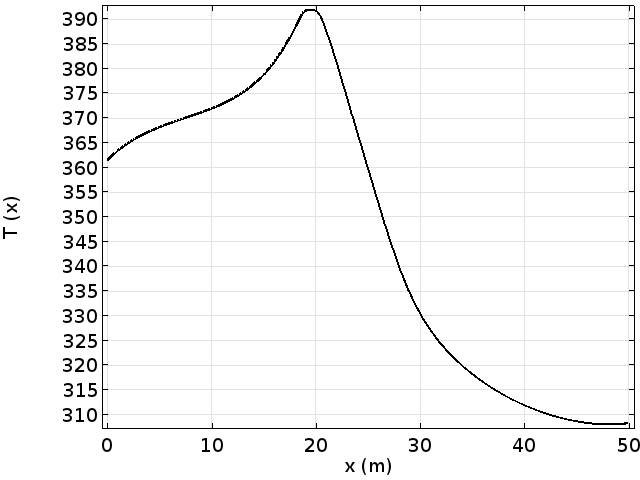}}\hfill
  \subfloat[]{\includegraphics[width=0.5\linewidth]{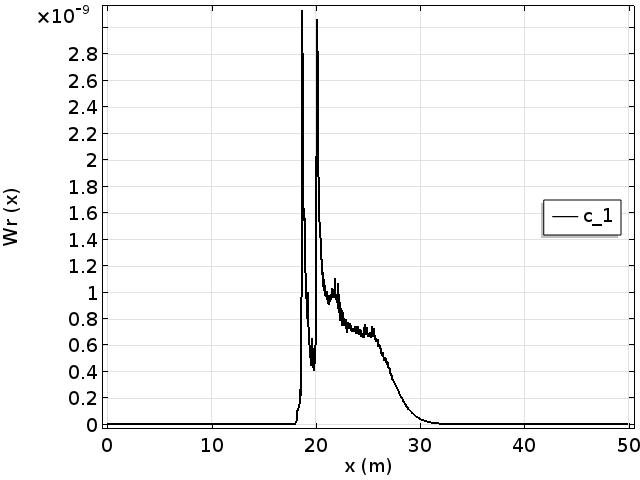}}\\
  \subfloat[]{\includegraphics[width=0.5\linewidth]{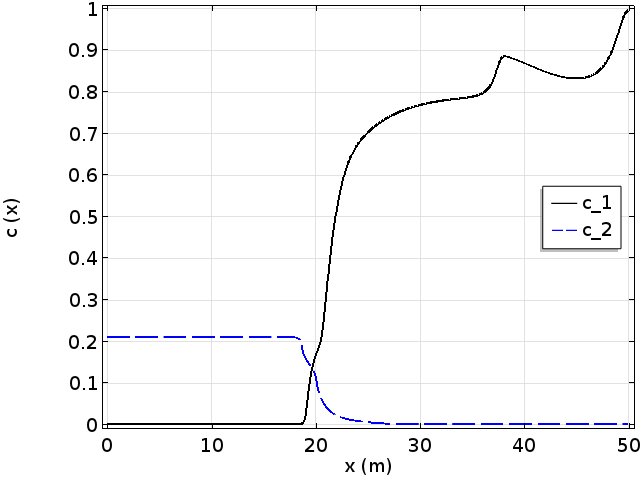}}
  \medskip
  \caption{Averaged values for (a) temperature, (b) reaction rate and (c) oil and oxygen fraction. Results are shown for $t=1.87\times10^8~s$}
  \label{figApp.average}
\end{figure}

In Fig. \ref{figApp.comp} we present the fractions at the simulation time of $t=1.87\times10^8~s$, which is when breakthrough of the inert components occurs.
We see that the inert components finger through the oil, but the oxygen is completely consumed by the reaction such that breakthrough of oxygen does not occur.
The localised aspect of combustion is best seen in Fig. \ref{figApp.temp}, where we present plots for the temperature field and the reaction $R$, as given by Eq. (\ref{eq.reac}).
The considerable lower value of combustion temperature is due to the additional thermal diffusion occurring in the vertical direction.
For the considered conditions, two fingers are formed and travel through the domain from injection to extraction point, i.e., from left to right of the domain.
From Figs \ref{figApp.comp}(c) and \ref{figApp.temp}(b) we can see that the reaction rate decreases along the fingers, as less oxygen is present.

In order to compare qualitatively the $2D$ results presented here with the results previously developed in this paper with the Koval model, we consider the integration of variables along the $y$-coordinate, as
\begin{equation}
    \overline{\varphi}(x) = \frac{\int_{y_{min}}^{y_{max}} \varphi(x,y) dy}{\int_{y_{min}}^{y_{max}} dy},
    \label{eqApp.int}
\end{equation}
where $y_{min}=0~m$ and $y_{max}=10~m$.
Then, the averaged values of temperature, reaction rate and fractions of oil and oxygen are shown in Fig. \ref{figApp.average} for $t=1.87\times10^8~s$.
From Fig. \ref{figApp.average}(b) it is possible to see that the reaction is fairly localised, which renders oxygen consumption to occur in a relatively thin region, as seen in Fig. \ref{figApp.average}(c).
The $1D$ description obviously fails to capture the fingering effects, which are relevant in the downstream side.
For the conditions considered in the $2D$ simulations, breakthrough of the inert components happened when the combustion wave starts to form, as seen by the small plateau around $x=20~m$ in Fig. \ref{figApp.average}(a).

\bibliographystyle{ieeetr}
\bibliography{refs.bib}

\end{document}